\documentclass[epjCONF]{svjour}
\usepackage{graphicx}
\usepackage[varg]{txfonts} 
\usepackage[latin1]{inputenc}
\session-title{%
19$^{\textnormal{\footnotesize th}}$ International %
IUPAP Conference on Few-Body Problems in Physics%
}
\begin{document}
\title{%
A Relativistic Coupled-Channel Formalism for the Pion Form Factor
}%
\author{%
E.P. Biernat\inst{1}\fnmsep\thanks{\email{ elmar.biernat@uni-graz.at} }
\and %
W. Schweiger\inst{1}
\and %
K. Fuchsberger\inst{2}
\and %
W.H. Klink\inst{3}
}
\institute{%
Institut f\"ur Physik, Universit\"at Graz, A-8010 Graz, Austria
\and %
BE-OP Division, CERN, CH-1211 Geneve 23, Switzerland
\and %
Department of Physics and Astronomy, University of Iowa, Iowa City, Iowa, USA
}
\abstract{ The electromagnetic form factor of a confined
quark-antiquark pair is calculated within the framework of
 point-form relativistic quantum mechanics. The dynamics of the
exchanged photon is explicitly taken into account by treating the
electromagnetic scattering of an electron by a meson as a
relativistic two-channel problem for a Bakamjian-Thomas type mass
operator. This approach guarantees Poincar\'e invariance. Using a
Fesh\-bach reduction the coupled-channel problem can be converted
into a one-channel problem for the elastic electron-meson channel.
By comparing the one-photon-exchange optical potential at the
constituent and hadronic levels, we are able to unambiguously 
 identify the electromagnetic meson form factor. Violations of
cluster-separability properties, which are inherent in the
Bakamjian-Thomas approach, become negligible for  sufficiently
large invariant mass of the electron-meson system. In the limit of an
infinitely large invariant mass, an equivalence with form-factor
calculations done in front-form relativistic quantum mechanics
is established analytically.
} 
\maketitle
%
%
%
\section{Introduction}
\label{BiernatEP_intro}
The invariant electron-hadron scattering amplitude in the
one-photon-exchange approximation can be written as the
contraction of a (point-like) electron current with a hadron
current times the photon propagator. The hadron current is a sum
of independent Lorentz covariants, which transform like
four-vectors and are multiplied by Lorentz invariant functions,
the hadron form factors. These form factors, which are functions
of the four-momentum-transfer squared ($q^2=-Q^2$), are the
observables that describe the electromagnetic structure of the
hadron. The present calculation of meson form factors is based on
the point form of relativistic quantum mechanics. This form is
characterized by the property that all four generators of
space-time translations are interaction dependent, whereas the
generators of Lorentz transformations stay free of interactions.

Our strategy is to start with a Poincar\'{e} invariant treatment
of electron-meson scattering. To this aim we adopt a two-channel
version of the Bakamjian-Thomas (BT) approach~\cite{BT53,KP91} and
work within a velocity-state representation~\cite{Kl98}, which is
particularly convenient for our purposes. The second channel
contains the photon. Its coupling to electrons, hadrons and quarks
can be taken from conventional quantum electrodynamics~\cite{Kl03}, but
has to be appropriately adapted to be applicable within the
BT-framework. In the following we will outline how this formalism
works at the hadronic level for the scattering of an electron by a
(spatially extended) pseudoscalar meson. Then we will proceed in
an analogous way at the constituent level. Observing that the
one-photon-exchange optical potential at both the hadronic and
the constituent levels have the structure of a cur\-rent-current
interaction then makes it possible to extract the meson current and,
consequently, the meson form factor.

\section{One-Photon-Exchange Optical Potential}
\label{BiernatEP_sec:1}
If the dynamics of the exchanged photon is to be taken into
account explicitly, the mass operator for electromagnetic
scattering of an electron by a pseudoscalar meson should act on a
direct sum of Hilbert spaces that accommodate the initial and
final electron-meson states as well as the intermediate
electron-meson-photon states. By projecting the mass eigenstates
onto these subspaces, the eigenvalue equation for the mass
operator becomes a system of two coupled equations for the
respective components. Elimination of the component containing the
photon then leads to a single equation for the electron-meson
component $\vert \psi_{e M} \rangle$:
\begin{eqnarray}
\label{eq:DynamicalEquationM} \left(\hat{M}_{e M }-m\right)\vert
\psi_{e M} \rangle &=& \hat{K}^\dag { \left(\hat{M}_{e M \gamma}
-m\right) }^{-1} \hat{K} \vert \psi_{e M} \rangle\nonumber\\ &=:& \hat{
V}_\mathrm{opt}(m)\, \vert \psi_{e M} \rangle\, .
\end{eqnarray}
Here $m$ is the mass eigenvalue, $\hat{K}$ and $\hat{K}^\dag$ are
vertex operators responsible for the emission and absorption of a
photon and $\hat{M}_{e M}$ and $\hat{M}_{e M \gamma}$ are mass
operators for the free electron-meson and electron-meson-photon
systems, res\-pectively. The right-hand side of
Eq.~(\ref{eq:DynamicalEquationM}) describes the action of the
one-photon-exchange optical potential $\hat{V}_\mathrm{opt}$ on
the $ \vert \psi_{e M} \rangle$ state. To extract the
meson current we calculate matrix elements of the optical
potential $\hat{V}_\mathrm{opt}$ in a velocity state
basis~\cite{Kl98}. On the hadronic level the structure of the
meson is encoded in a phenomenological form factor $f(Q)$ which is
inserted by hand at the photon-meson vertex. Furthermore, in order
to be able to work within the BT framework~\cite{BT53} one has to
resort to the approximation that the four-velocity of the whole
($eM-eM\gamma$) system is conserved at the electromagnetic
vertices. With these assumptions and neglecting self-energy
contributions due to photons being emitted and absorbed by the
same particle the on-shell matrix elements of the optical
potential on the hadronic level read~\cite{Fu03}
\begin{eqnarray}\label{eq:1phexchamp}
\lefteqn{\langle v^\prime; \vec{k}_e^\prime, \mu_e^\prime;
\vec{k}_M^\prime \vert\, \hat{ V}_\mathrm{opt}(m)\, \vert v;
\vec{k}_e, \mu_e; \vec{k}_M \rangle_{\mathrm{on-shell}}}\nonumber\\
&\propto& v_0\, \delta^3 (\vec{v}^{\, \prime} - \vec{v}\, )
\,f(Q)\, j_\mu(\vec{k}_M^\prime ;
\vec{k}_M)
\,
\frac{(-\mathrm g^{\mu\nu})}{Q^2}\;j_\nu(\vec{k}_e^\prime,
\mu_e^\prime;\vec{k}_e, \mu_e) \, .\nonumber\\
\end{eqnarray}
Here $v^{(\prime)}$ is the incoming (outgoing) total velocity of
the electron-meson system, $\vec{k}_{e}^{(\prime)}$ and
$\vec{k}_{M}^{(\prime)} = -\vec{k}_{e}^{(\prime)}$ are the
incoming (outgoing) center-of-mass momenta of the electron and
meson, respectively. \lq\lq On-shell\rq\rq\ means that due to the
center-of-mass kinematics and the fact that we are only interested
in the one-photon-exchange amplitude we can set
$m=(m_e^2+\vec{k}_{e}^2)^{1/2}+(m_M^2+\vec{k}_{M}^2)^{1/2}$ and
$\vert\vec{k}_{M}\vert = \vert\vec{k}_{M}^{(\prime)}\vert
=\vert\vec{k}_{e}\vert =\vert\vec{k}_{e}^{(\prime)}\vert$. The
$\mu_e$'s are the spin-projections of the electron and
$j_\nu(\vec{k}_e^\prime, \mu_e^\prime;\vec{k}_e, \mu_e)$ as well
as $j_\mu(\vec{k}_M^\prime ; \vec{k}_M)$ are the point-like
electromagnetic electron and meson currents, respectively. Apart
from a kinematical factor (which has been dropped for better
readability) the right-hand side of Eq.~(\ref{eq:1phexchamp})
corresponds to the familiar one-photon exchange amplitude for
elastic electron-meson scattering (calculated in the
center-of-mass system).

In order to achieve a microscopic description of the form factor
$f(Q)$ we employ a constituent-quark model in which the meson is
composed of a quark and an antiquark. Working within the same
approach as outlined above, the mass operator has now to be
defined on a Hilbert space that comprises electron-quark-antiquark
and electron-quark-an\-ti\-quark-photon states. Furthermore we
assume that an instantaneous confinement potential acts between
quark and an\-ti\-quark (in both channels) and, as before, the
total velocity should be conserved at the photon-electron and
photon-(anti)quark vertices. For a comparison with
Eq.~(\ref{eq:1phexchamp}) one has to calculate (on-shell) matrix
elements of the optical potential on the constituent-level,
$\hat{V}_\mathrm{opt}^{\mathrm{const}}$, between quark-antiquark
bound states with the quantum numbers of the meson. Neglecting
again self-energy contributions a tedious calculation yields
~\cite{Fu03}:
\begin{eqnarray}\label{eq:1phexchampconst}
\lefteqn{\langle v^\prime; \vec{k}_e^\prime, \mu_e^\prime;
\vec{k}_M^\prime \vert\, \hat{ V}_\mathrm{opt}^{\mathrm{const}}(m)\, \vert v;
\vec{k}_e, \mu_e; \vec{k}_M \rangle_{\mathrm{on-shell}}}\nonumber\\
&\propto& v_0\, \delta^3 (\vec{v}^{\, \prime} - \vec{v}\, )\,J_\mu(\vec{k}_M^\prime ;
\vec{k}_M)
\,
\frac{(-\mathrm g^{\mu\nu})}{Q^2}\; j_\nu(\vec{k}_e^\prime, \mu_e^\prime;\vec{k}_e, \mu_e)  \,
.\nonumber\\
\end{eqnarray}
Here we have dropped the same kinematical factor as in
Eq.~(\ref{eq:1phexchamp}). The microscopic meson current
$J_\nu(\vec{k}_M^\prime ; \vec{k}_M)$ is a rather lengthy
expression of an integral over bound state wave functions, quark
currents and Wigner $D$ functions~\cite{Bi09}.

\section{Meson Form Factor}
\label{BiernatEP_sec:2}
If the quark-antiquark bound state at the hadronic level has the same
 quantum numbers as the meson, we can equate the right-hand
sides of Eqs.~(\ref{eq:1phexchamp}) and~(\ref{eq:1phexchampconst})
to find an expression for the form factor~\cite{Fu03}
\begin{eqnarray}\label{eq:ff}
 f(Q,\vert\vec{k}_M\vert)=\frac{j^\mu(\vec{k}_e^\prime, \mu_e^\prime;\vec{k}_e, \mu_e)
 \;  J_\mu(\vec{k}_M^\prime ;
\vec{k}_M)}{j^\nu(\vec{k}_e^\prime, \mu_e^\prime;\vec{k}_e, \mu_e) \;
j_\nu(\vec{k}_M^\prime ;
\vec{k}_M)}\,.
\end{eqnarray}
Here we have introduced the magnitude of the center-of-mass
momentum of the meson $\vert\vec{k}_M\vert$ as a further argument
of the vertex form factor. This additional dependence of the form
factor on $\vert\vec{k}_M\vert$ (or equivalently the invariant
mass of the electron-meson system) is a consequence of 
velocity conservation at the electromagnetic vertices, which is required
 to get a BT-type mass operator. It should be emphasized that such an additional
$\vert\vec{k}_M\vert$-dependence of the meson form factor still
preserves Poincar\'e invariance. But it violates cluster
separability. Violation of cluster separability, in particular the
violation of the cluster condition for the Poincar\'{e}
generators, is nearly unavoidable within the BT-framework, but can
be overcome by the introduction of, so called, \lq\lq packing
operators\rq\rq~\cite{KP91}. \footnote{It should be noted that for
the case of an interacting pair and a spectator our
velocity-state representation of operators \textit{does not}
violate  cluster separability for the scattering
operator~\cite{Co65}. The separability condition for the scattering operator
should, however, be contrasted with the stronger condition for the
generators~\cite{CP82}.} Fortunately this
$\vert\vec{k}_M\vert$-dependence vanishes rather quickly with
increasing $\vert\vec{k}_M\vert$ and suggests
taking the limit $\vert\vec{k}_M\vert \rightarrow\infty$. In this
limit the microscopic meson current factorizes explicitly into a
point-like meson current times a form factor:
\begin{eqnarray}\label{eq:microcurrentlim}
 J_\nu(\vec{k}_M^\prime ;
\vec{k}_M)\stackrel{\vert\vec{k}_M\vert \rightarrow\infty }
{\longrightarrow} F(Q)\, j_\nu(\vec{k}_M^\prime ;
\vec{k}_M)\,.\end{eqnarray} The final result for the form factor
has a simple analytical form ~\cite{Bi09}:
\begin{eqnarray}\label{eq:finalresult}
F(Q)=\int\mathrm d^3 \tilde k_q^\prime \,\sqrt{\frac{m_{q\bar q}}
{m_{q\bar q}^\prime}}\,\mathcal S\,
\Psi^{\ast}\left(\tilde{\vec{k}}^\prime_q\right)\,
\Psi\left(\tilde{\vec{k}}_q\right)\,.
 \end{eqnarray}
Here $Q^2=\vec q^2$ is the momentum transfer squared with
$\vec{q}=\vec{k}_q'-\vec{k}_q=\vec {k}_M'-\vec{k}_M$ and $m_{q\bar
q}^2=(E_q+E_{\bar q})^2-\vec{k}_M^2$ is the invariant mass of the
quark-antiquark pair. Quantities without a tilde refer to the
electron-meson center-of-mass and quantities with a tilde to the
meson rest system. $\Psi$ is the bound-state wave function of the quark-antiquark pair and $\mathcal{S}$ is a spin-rotation factor which
takes into account the substantial effect of the quark spin on the
form factor. By an appropriate change of variables the integral
for the form factor, Eq.~(\ref{eq:finalresult}), takes the same
form as the integral for the pion form factor from front-form
calculations~\cite{CC88,CP05}. This remarkable result means that
relativity is treated in an equivalent way and the physical
ingredients are the same in both approaches.

\begin{figure}[!h]
\centering
\includegraphics[width=1\columnwidth,angle=0]{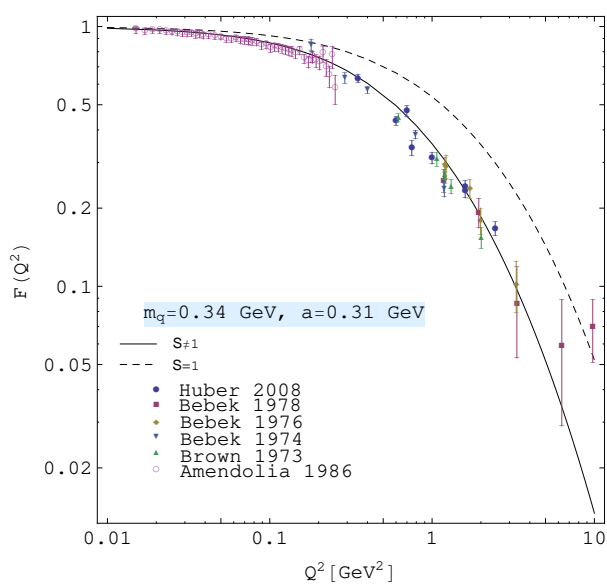}
\caption{\label{fig:1}\small $Q^2$-dependence of the pion form
factor with (solid) and without (dashed) spin-rotation factor
$\mathcal S$. Values for the quark mass $m_q$ and the oscillator
parameter $a$ are taken from~~\cite{CC88,CP05} and data are taken
from~\cite{Am86,Br73,Be74,Be76,Be78,Hu08}.}
\end{figure}

For a simple two-parameter harmonic-oscillator wave function with
the parametrization taken from~\cite{CC88,CP05} our result for the
pion electromagnetic
 form factor
provide a reasonable fit to the data as shown in Fig.~\ref{fig:1}.
\section{Summary and Outlook}
\label{BiernatEP_sec:4}
We have applied the point form of relativistic quantum mechanics
in connection with the BT-formalism to analyze the electromagnetic
structure of pseudoscalar mesons. In our approach the scattering
of an electron by a confined quark-antiquark pair is treated as a
two-channel problem for the mass operator. In this way the
dynamics of the exchanged photon can be taken into account
explicitly. Quark confinement is treated via an instantaneous potential.
The emission and absorption of a photon by an electron or
(anti-)quark is described by a vertex interaction that has the
Lorentz structure of the field theoretical vertex but conserves
the four-velocity of the whole system. By construction this
approach is Poincar\'{e} invariant. We are then able to identify
the meson form factor from the one-photon-exchange optical
potential in an unambiguous way. The extracted meson form factor,
however,  depends not only on Mandelstam $t=-Q^2$ (the
four-momentum transfer squared) but also shows a (mild) dependence
on Mandelstam $s$ (the total invariant mass squared of the
$eM$-system). This additional $s$-dependence does not spoil
Poin\-car\'{e} invariance but indicates a violation of cluster
separability. It is a consequence of working with the point-form
version of the BT-formalism which demands velocity conservation at
each interaction vertex. The observation that the $s$-dependence
vanishes quickly with increasing $s$ indicates, however, that
cluster-separability-violating effects are negligible for
sufficiently large $s$.
\footnote{In point form it seems  natural to use Lorentz boosts in
opposite directions  to space-like separate
subsystems~\cite{BKS2010}. In this way one stays on the
quantization hypersurface. Separations by boosts, however,
increases the invariant mass of the whole system. Or, reversing the
argument, by taking $s$ large we separate the electron from the
meson such that it does not affect the meson structure.}
Indeed, as we show analytically, in the limit of infinitely
large $s$ the microscopic meson current goes over into a product
of the usual point-like meson current times an integral with the
integrand depending only on the momentum transfer $Q$ and on
internal variables (that are integrated and summed over). In this
limit the form factor acquires a simple analytical form which has
been shown to be equivalent to the usual front-form expression in
a $q^+ = 0$ frame.

Our multichannel approach for the calculation of form factors can
easily be generalized to electroweak form factors of
arbitrary few-body bound systems. By an appropriate extension of
the Hilbert space it should also be possible to handle
exchange-current effects within this type of approach.

\end{document}